\documentclass[a4paper,fleqn,usenatbib]{mnras}

\usepackage{newtxtext,newtxmath}

\usepackage[T1]{fontenc}
\usepackage{ae,aecompl}

\usepackage{graphicx}	
\usepackage{amsmath}	
\usepackage{amssymb}	
\usepackage{multicol}        
\usepackage{bm}		
\usepackage{pdflscape}	

\usepackage{epsfig}
\usepackage{threeparttable} 
\usepackage{rotating}

\usepackage{epstopdf}
\usepackage[usenames, dvipsnames]{color}

\usepackage{booktabs,array}

\usepackage{color}

\newcommand{\MAXI}{MAXI~J1813--095}

\newcommand{\ha}{H$\alpha$}

\newcommand{\lx}{$L_\mathrm{X}$}

\newcommand{\Nh}{$N_{\rm H}$}
\newcommand{\qhired}{$\chi^{2}_{\nu}$}
\newcommand{\qhis}{$\chi^{2}$}

\newcommand{\Msun}{\mathrm{M}_{\odot}}
\newcommand{\lum}{\mathrm{erg~s}^{-1}}
\newcommand{\flux}{\mathrm{erg~cm}^{-2}~\mathrm{s}^{-1}}
\newcommand{\cnts}{\mathrm{counts~s}^{-1}}
\newcommand{\nh}{$\mathrm{cm}^{-2}$}

\newcommand{\kms}{\mbox{${\rm km}\:{\rm s}^{-1}\:$}}

\newcommand{\swift}{\textit{Swift}}
\newcommand{\xmm}{\textit{XMM-Newton}}

\newcommand{\inte}{\textit{INTEGRAL}}

\newcommand{\maxi}{\textit{MAXI}}

\title[The black hole candidate MAXI~J1813--095]{Multi-wavelength spectroscopy of the black hole candidate MAXI~J1813--095 during its discovery outburst}

\author[M. Armas Padilla]{M. Armas Padilla,$^{1,2}$\thanks{E-mail: m.armaspadilla@iac.es} 
T. Mu\~noz-Darias,$^{1,2}$
J. S\'anchez-Sierras,$^{2}$
B. De Marco,$^{3}$
\newauthor 
F. Jim\'enez-Ibarra,$^{1,2}$
J. Casares,$^{1,2}$
J. M. Corral-Santana,$^{4}$
and M. A. P. Torres $^{1,2,5}$ 
\\
$^{1}$Instituto de Astrof\'isica de Canarias (IAC), V\'ia L\'actea s/n, La Laguna 38205, S/C de Tenerife, Spain\\
$^{2}$Departamento de Astrof\'isica, Universidad de La Laguna, La Laguna, E-38205, S/C de Tenerife, Spain\\
$^{3}$Nicolaus Copernicus Astronomical Center, Polish Academy of Sciences, Bartycka 18, PL-00-716 Warsaw, Poland\\
$^{4}$European Southern Observatory (ESO), Alonso de C\'ordova 3107, Vitacura, Casilla 19, Santiago, Chile\\
$^{5}$SRON, Netherlands Institute for Space Research, Sorbonnelaan 2, 3584~CA, Utrecht, The Netherlands
}

\date{Accepted XXX. Received YYY; in original form ZZZ}

\pubyear{2019}

\begin{document}
\label{firstpage}
\pagerange{\pageref{firstpage}--\pageref{lastpage}}
\maketitle

\begin{abstract}
\MAXI\ is an X-ray transient discovered during an outburst in 2018. We report on X-ray and optical observations obtained during this event, which indicate that the source is a new low-mass X-ray binary. The outburst lasted $\sim$70~d and peaked at \lx(0.5--10keV)$\sim7.6\times10^{36}~\lum$, assuming a distance of 8 kpc. \swift/XRT follow-up covering the whole activity period shows that the X-ray emission was always dominated by a hard power-law component with a photon index in the range of 1.4--1.7. These values are consistent with \MAXI\ being in the hard state, in agreement with the $\sim$30 per cent fractional root-mean-square amplitude of the fast variability (0.1--50~Hz) inferred from the only \xmm\ observation available. The X-ray spectra are well described by a Comptonization emission component plus a soft, thermal component ($kT\sim$0.2~keV), which barely contributes to the total flux ($\lesssim$8~per cent). The Comptonization y-parameter ($\sim$1.5), together with the low temperature and small contribution of the soft component supports a black hole accretor. We also performed optical spectroscopy using the VLT and GTC telescopes during outburst and quiescence, respectively. In both cases the spectrum lack emission lines typical of X-ray binaries in outburst. Instead, we detect the \ion{Ca}{ii} triplet and \ha\ in absorption. The absence of velocity shifts between the two epochs, as well as the evolution of the \ha\ equivalent width, strongly suggest that the optical emission is dominated by an interloper, likely a G--K star. This favours a distance $\gtrsim$3~kpc for the X-ray transient.
\end{abstract}




\section{Introduction}\label{sec:intr}

The most extreme stellar compact objects, black holes (BHs) and neutron stars (NSs), are often revealed when they reside in low mass X-ray binary (LMXB) systems. In these interacting binaries, the compact object is accreting material from the outer layers of a (sub-)solar companion star that overflows its Roche lobe. Due to conservation of angular momentum, the in-falling gas forms a disc around the compact object, an accretion disc \citep{Shakura1973}, which can exceed $10^{6-7}$~K in its innermost regions, emitting strong X-ray radiation. Depending on their X-ray luminosity (a proxy of the accretion rate) LMXBs behave as persistent or transient sources. The former are persistently accreting and, consequently, they are always luminous, with \lx$>10^{36}~\lum$ (but see \citealt{ArmasPadilla2013b}). The latter, on the other hand, combine long quiescent states at low X-ray luminosities (\lx$\sim10^{30-33}~\lum$) with sporadic outburst events, during which they brighten several orders of magnitude. Over these periods of activity (lasting weeks to months), LMXBs display a well established X-ray phenomenology, exhibiting different X-ray states in which their spectral and timing properties vary  following (roughly) the changes in the accretion rate \citep{Miyamoto1992,Homan2001,Klis2006,Belloni2011}. In the optical/infrared, active LMXBs are characterised by broad, double-peaked emission lines arising from hot gas orbiting in a Keplerian accretion disc geometry \citep{Smak1969}. These features can sometimes show very complex profiles due to e.g. the presence of outflows \citep{Munoz-Darias2016} or changes in the disc opacity \citep{Jimenez-Ibarra2019,Dubus2001}. Finally, ejection processes in the form of jets are ubiquitously  observed in the low energy part of the spectrum, from radio waves to (at least) the near-infrared \citep[e.g.][]{Fender2004}.

Despite of this wealth of information, it is not always straightforward to infer the nature of the compact object since most of the above phenomenology is common to BH and NS systems \citep{Fender2016,Charles2006}. We can securely establish the BH nature when the dynamical mass of the compact object exceeds  $\sim$3~$\Msun$ \citep{Casares2014}, the absolute upper limit for NSs according to General Relativity \citep{Rhoades1974, Kalogera1996}. NS systems, on the other hand, are unveiled when events associated to the presence of a hard surface are witnessed. These are thermonuclear X-ray bursts and coherent pulsations. In the absence of any of the above, one relies on either the long-term evolution or a very detailed analysis of the X-ray timing and spectral characteristics. From the spectral point of view,  a thermal accretion disc and a hard tail produced by inverse-Compton processes in an optically thin inner flow (corona) are common X-ray spectral components for both NSs and BHs. In addition, NSs possess an extra soft thermal component that arises from the solid surface and/or boundary layer. This component is clearly detected in high signal to noise data of some NS systems (e.g. \citealt{ArmasPadilla2017}). In addition, the latter it is a source of extra seed photons affecting the properties of the Comptonizing Corona, which can be used to constrain the nature of the accretor \citep{Sunyaev1989,Burke2017}.

On 2018 February 19 the Monitor of All-sky X-ray Image (\maxi) reported the detection of an uncatalogued X-ray transient, \MAXI\ \citep{Kawase2018}. Following its discovery, the source was monitored by several observatories. Quasi-simultaneous spectra obtained with \swift\ and \inte\ missions were well modelled by an absorbed powerlaw with photon--index $\Gamma$=1.6 and cutoff energy at $\sim$140~keV \citep{Fuerst2018}. The Australia Telescope Compact Array detected the source with a flux density of $\sim$0.69~mJy at 5.5 GHz and $\sim$0.47~mJy at 9 GHz. This results in a radio spectral index of $\alpha\sim-0.6$, which tentatively points towards a compact jet from a BH in the hard-state \citep{Russell2018}. GROND \citep{Greiner2008}, attached to the MPG 2.2m telescope at the ESO La Silla Observatory (Chile), detected a candidate optical counterpart at i=18.6 [$\rmn{RA}(\rm J2000)=18^{\rmn{h}} 13^{\rmn{m}} 34\fs015$,
$\rmn{Dec.}~(\rm J2000)=-09\degr 31\arcmin 59\farcs20$] which had brightened by ~1 magnitude with respect to Pan--STARRS1 (PS1) pre-outburst observations \citep{Rau2018}.  

Here, we present a multi-wavelength study of the discovery outburst of \MAXI. This includes X-ray spectral and temporal analysis from \xmm\ and \swift\ as well as optical spectroscopy taken with the Very Large Telescope (VLT) during outburst and the Gran Telescopio Canarias (GTC) in quiescence.

\section{Observations and data reduction}\label{sec:obs}
\begin{figure}
\begin{center}
\includegraphics[keepaspectratio,width=\columnwidth, trim=0.0cm 0.0cm 0.0cm 0.0cm]{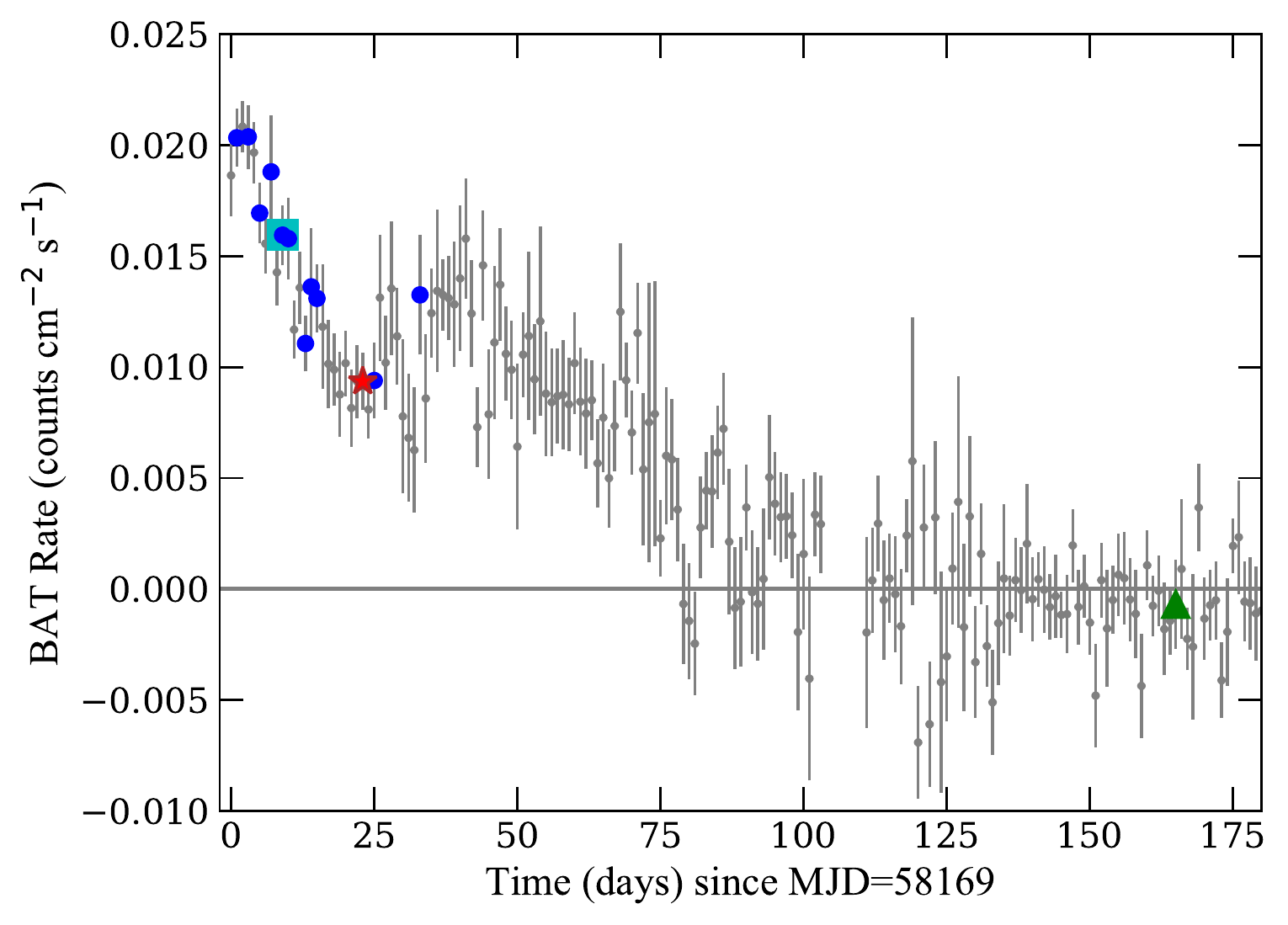}\\
\caption{The 2018 outburst light curve (15--50 keV) of \MAXI\ obtained by \swift/BAT. The  dates of the \swift/XRT and \xmm\ observations are indicated by blue circles and a red star, respectively. A cyan square (VLT) and a green triangle (GTC) indicate the epochs of our optical spectroscopy.}

\label{fig:LC}
\end{center}
\end{figure}


\subsection{\swift\ data}\label{subsec:swift}

The Neil Gehrels \swift\ Observatory \citep{Gehrels2004} pointed to \MAXI\ on 11 occasions with the on-board X-ray Telescope (XRT; \citealt{Burrows2005}) operated in window timing (WT) mode (see Fig \ref{fig:LC} and Table \ref{tab:swiftres} for the observing log).  We used the {\ttfamily HEASoft} v.6.20 package to reduce the data. We processed the raw XRT data running the {\ttfamily xrtpipeline} task selecting the standard event grades of 0-2. We extracted (using {\ttfamily Xselect}~v2.4) the source spectra, light curves and images from a circular region of $\sim$60~arcsec (25 pixels) radius. To compute the background, we used an annulus centred on the source with an 180~arcsec (75 pixels) inner radius and 300~arcsec (125 pixels) outer radius. We created exposure maps and ancillary response files following the standard \swift\ analysis threads\footnote{\url{http://www.swift.ac.uk/analysis/xrt/}}, and acquired the last version of the response matrix files from the High Energy Astrophysics Science Archive Research Center calibration database. In order to be able to use \qhis\ statistics consistently, the spectral data were grouped in bins containing a minimum of 20 photons. 

\subsection{\xmm\ data}\label{subsec:xmm}

\xmm\ \citep{Jansen2001} observed \MAXI\ during 26.7~ksec on 2018 March 15 (ID 0830190101). During the trigger, the European Photon Imaging Camera (EPIC) was operated with the thin filter in the small-window imaging mode, in the case of MOS1 camera, and the timing mode for MOS2 and PN cameras (\citealt{Turner2001},\citealt{Struder2001}). We used the Science Analysis Software (SAS, v.16.0.0) to obtain calibrated events and scientific products.

We excluded events with high background flaring activity by filtering the periods with count rates larger than 0.3~$\cnts$ (> 10~keV) and 0.7~$\cnts$ (10--12~keV) for the MOS1 and PN cameras, respectively. MOS1 source events have a count rate of $\sim17~\cnts$, well beyond the threshold of $\sim5~\cnts$ where pile up starts to be an issue in the small-window mode. We confirmed this by using the pile up evaluation tool EPATPLOT. In order to mitigate pile up effects we used an annulus region excluding a large central portion (15 arcsec inner radius and a 40 arcsec outer radius) to extract the source filtered events. This resulted in a low statistic spectrum and, therefore, we did not include MOS1 data in our study. We also excluded MOS2 data from our analysis because of the calibration uncertainties in the MOS cameras in the timing mode \footnote{See the calibration technical note XMM-SOC-CAL-TN-0018}.

We extracted PN source and background events from the RAWX columns in [30:44] and in [6:14], respectively. The source count rate is $\sim60~\cnts$, which is much lower than the 800~$\cnts$ pile-up threshold. We generate the light curves, spectra, associated response matrix files (RMFs) and ARFs following the standard analysis threads \footnote{\url{https://www.cosmos.esa.int/web/xmm-newton/sas-threads}}.  

The source was detected by the Reflection Grating Spectrometer (RGS) which was operated in the standard spectroscopy mode. We used the SAS task RGSPROC to reduce the RGS data and to produce the response matrices and spectra. We included the RGS1 and RGS2 first order spectra in our analysis.

We rebinned the spectra in order to include a minimum of 25 counts in every spectral channel, avoiding to oversample the full width at half-maximum of the energy resolution by a factor larger than 3.

\subsection{Optical observations}
 
We carried out optical spectroscopy of  \MAXI\ at two different epochs: close to the (observed) peak of the outburst and $\sim 150$~d later, when the source was no longer detected by X-ray monitors and was expected to be near its quiescent level (Fig. \ref{fig:LC}). Outburst observations were carried on March 1, 2018 using the instrument X-shooter \citep{Vernet2011} attached to the VLT-UT2 telescope in Cerro Paranal, Chile. This instrument has 3 different arms covering the ultraviolet, visible and near-infrared spectral ranges.  The source was not detected in the ultraviolet due to the strong reddening, while the low signal-to-noise achieved in the infrared prevented us from carrying out an adequate telluric correction. Therefore, we use only data from the visible X-Shooter arm in this paper. We obtained two exposures of 1205 seconds each, that were processed and combined using version~3.2.0 of the X-shooter pipeline.  Observations were obtained under clear conditions and a seeing of $\sim0.6$~arcsec, while a slit-width of 0.9~arcsec was used. This combined with a $2\times2$ binning rendered a velocity resolution of $\sim$ 35 km s$^{-1}$. 

We observed again \MAXI\ using the Optical System for Imaging and low-Intermediate Resolution Integrated Spectroscopy \citep[OSIRIS;][]{Cepa2000} attached to the GTC  at the Observatorio del Roque de Los Muchachos located in La Palma, Spain. Observations took place under clear conditions on August 3, 2018. We used the grism R2500R (1.04~\AA\ pix$^{-1}$), which covers the spectral range 5575--7685~\AA. We took one spectrum of 1800 s using a slit-width of 0.8~arcsec, which provided a spectral resolution of 137~km~s$^{-1}$ (measured as the full-width at half-maximum of a sky line at $\sim$ 6300~\AA). We reduced the data and extracted the spectrum using \textsc{iraf} standard routines and custom \textsc{python} software.  We used sky emission lines and the \textsc{molly} software to measure and correct subpixel velocity drifts of $\sim$24~km~s$^{-1}$ introduced by flexure effects. 

For both the VLT and GTC observations we obtained photometric measurements from the acquisition images. These were reduced in the standard way using \textsc{astropy-ccdproc} based routines \citep{Astropy2013,Astropy2018}. Aperture photometry was calibrated against nearby stars present in the PS1 catalog \citep{PanStarrs}. We obtained i-band magnitudes of $18.62 \pm 0.02$ and $19.43 \pm 0.03$, respectively. The former is consistent with the outburst value reported by \citet{Rau2018}, while the latter is slightly brighter than the pre-outburst PS1 magnitude ($i=19.67 \pm 0.03$).
 
\section{Analysis and results}\label{sec:anares}

\begin{figure}
\begin{center}
\includegraphics[keepaspectratio,width=\columnwidth, trim=0.0cm 0.0cm 0.0cm 0.0cm]{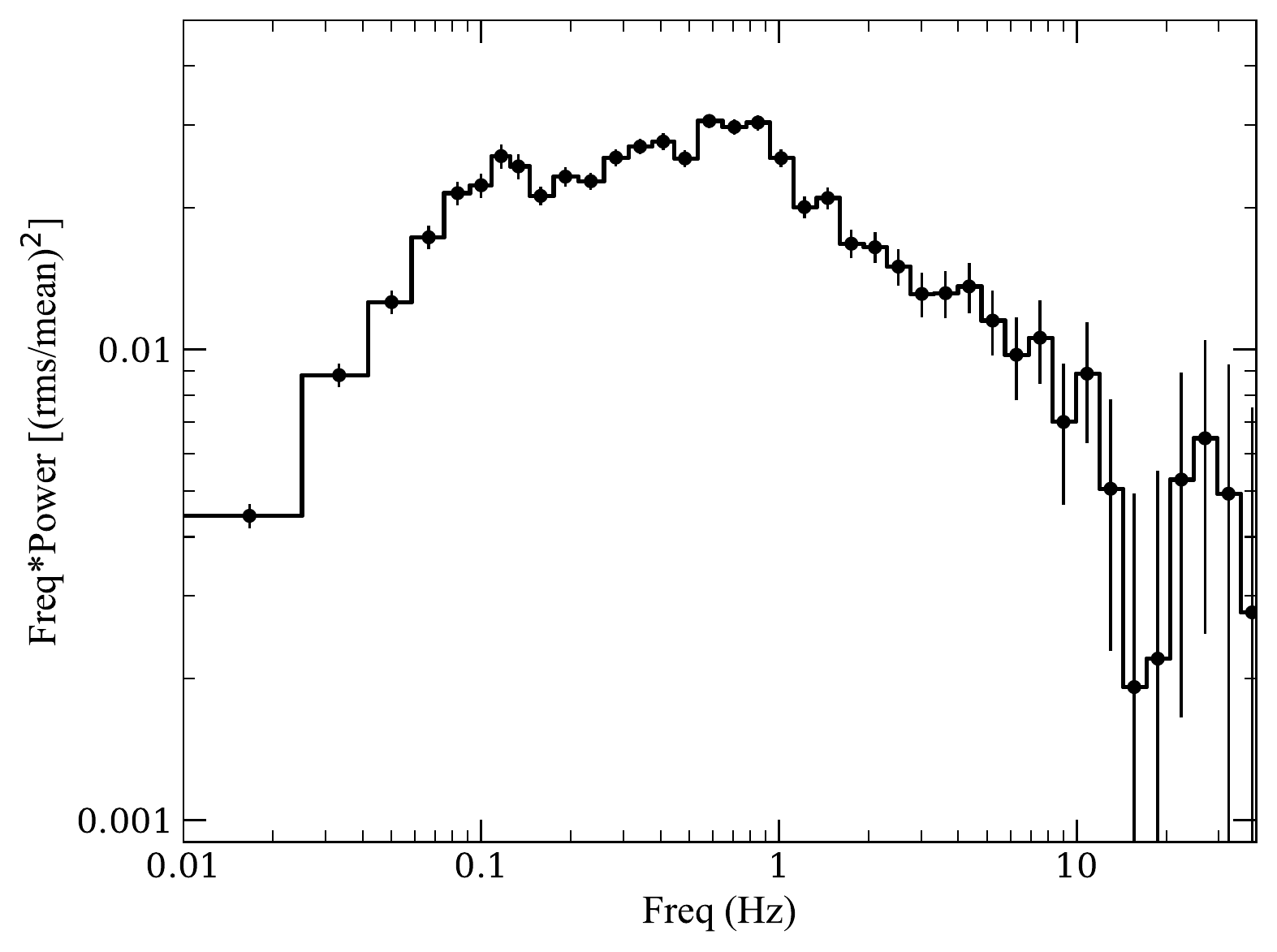}\\
\caption{The PSD of \MAXI\ in the 1--10~keV energy band. It is displayed Poisson noise-subtracted and using the fractional root-mean-square normalization \citep{Miyamoto1991}.}
\label{fig:pow_spec}
\end{center}
\end{figure}


\begin{figure}
\begin{center}
\includegraphics[keepaspectratio,width=\columnwidth, trim=0.0cm 0.0cm 0.0cm 0.0cm]{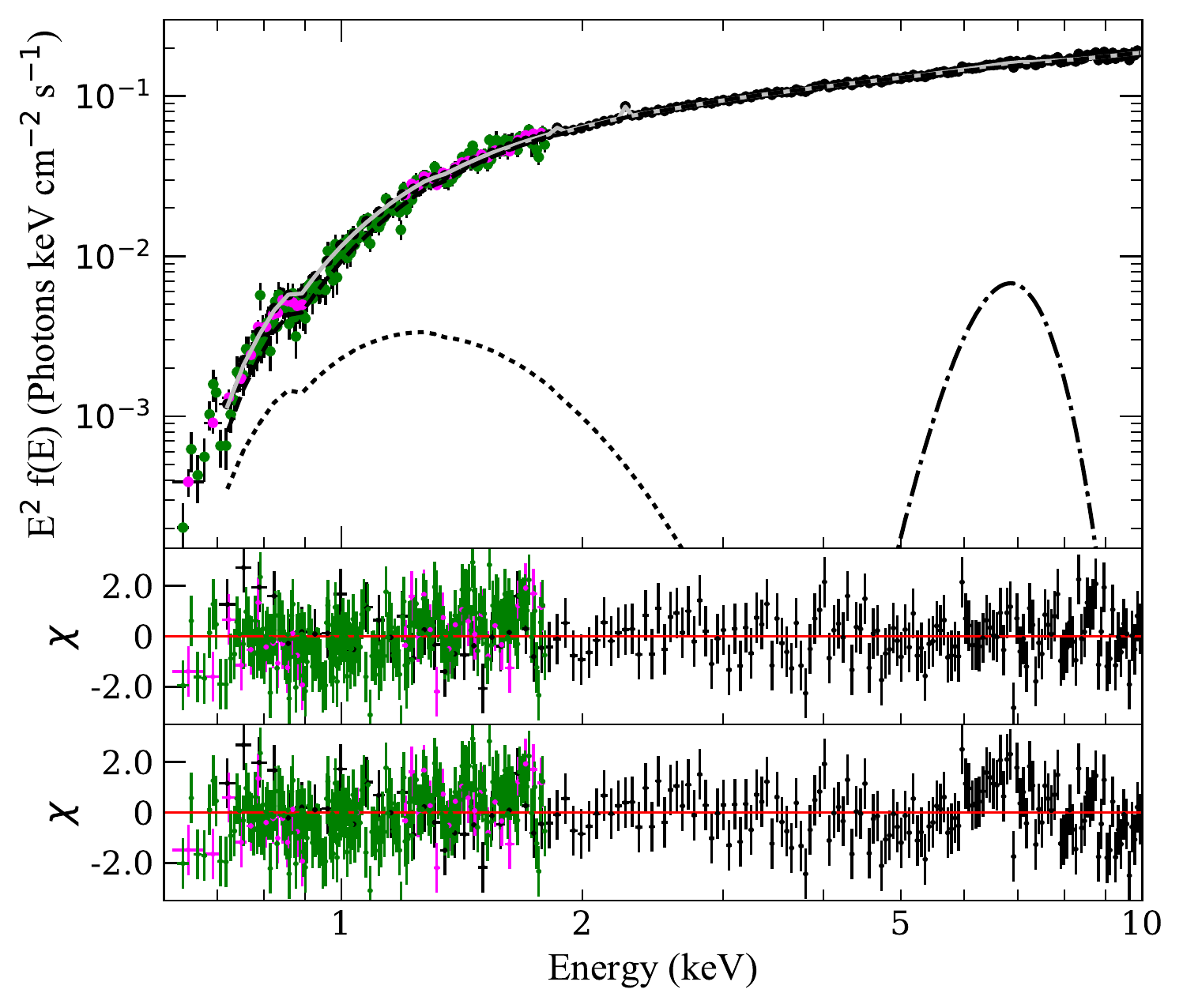}\\
\caption{ Unfolded PN (black), RGS1 (magenta) and RGS2 (green) spectra and residuals in units of $\sigma$ when using a DISKBBODY+NTHCOMP+GAUSS model. The full fit is shown as a grey solid line, the black body component as a dotted line, the Comptonized component as a dashed line and the Gaussian component (associated to the Fe K$\alpha$ line) as a dot-dashed line. Botton panel: Residuals in units of $\sigma$ for a DISKBBODY+NTHCOMP model (i.e., no Gaussian component).}

\label{fig:spectra}
\end{center}
\end{figure}


\begin{figure}
\begin{center}
\includegraphics[keepaspectratio,width=\columnwidth, trim=0.0cm 0.0cm 0.0cm 0.0cm]{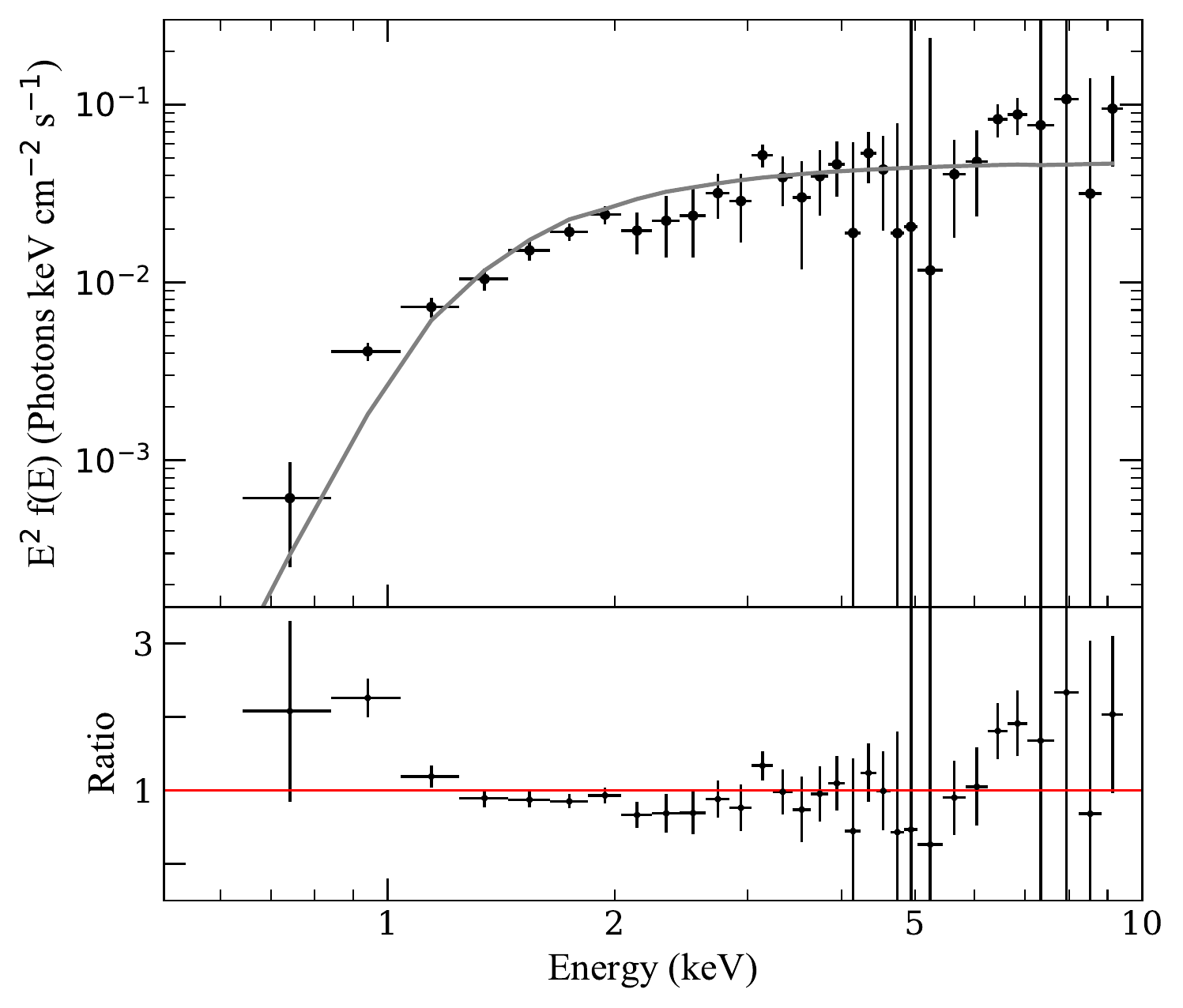}\\
\caption{Upper panel: rms-scaled energy spectrum (0.1--50~Hz) derived from the \xmm\ data. Bottom panel: Data-to-model ratio obtained by fitting the rms-scaled spectrum with TBABS*NTHCOMP model (grey solid line).}
\label{fig:RMS_spec}
\end{center}
\end{figure}


\begin{figure}
\begin{center}
\includegraphics[keepaspectratio,width=\columnwidth, trim=0.0cm 0.0cm 0.0cm 0.0cm]{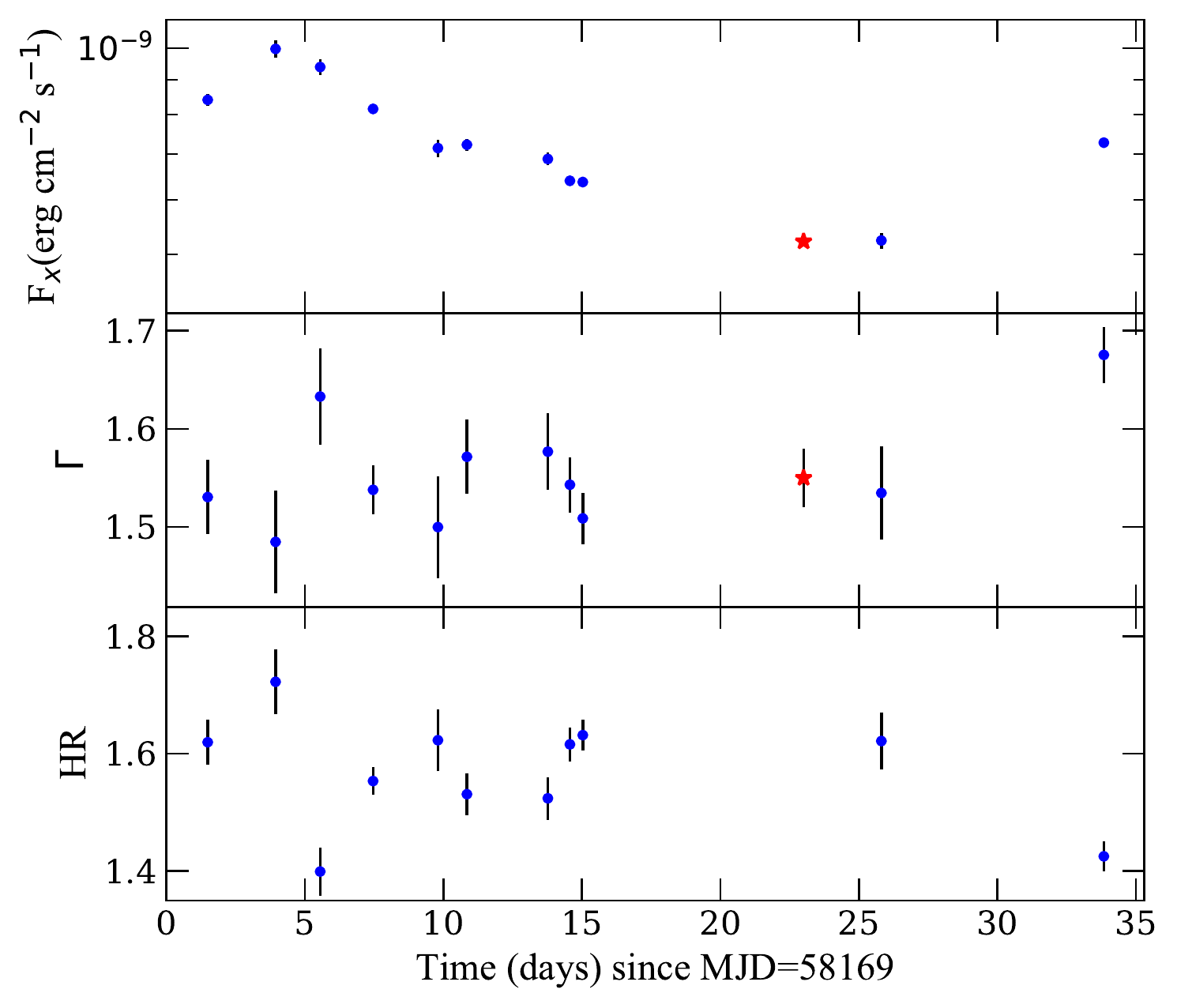}\\
\caption{From top to bottom, evolution with time of unabsorbed flux, photon index and hardness ratio (ratio of counts in the 2--10~keV and 0.5--2~keV energy bands). Fluxes are  computed in the 0.5--10~keV band. \swift\ data are indicated by a blue circle and \xmm\ observation by a red star.}

\label{fig:swift}
\end{center}
\end{figure}


%
\begin{table*}
\caption{Results from \swift\ spectral fits with an absorbed power-law model in which \Nh\ was fixed to 1.1$\times10^{22}$ \nh. Quoted errors represent 90\% confidence levels.}
\begin{threeparttable}
\begin{tabular}{l c c c c c c c c}
\hline \hline
Obs ID 		&Date &Exposure(ks)  & $\Gamma$  		& $F_{\rm X, unabs}^{\rm a}$ &  \lx$^{\rm b}$ & $F_{\rm X, unabs}^{\rm a}$ 	& \lx$^{\rm b}$		& \qhired\ (dof)  \\
 			& 	 		& 		 &   				&  \multicolumn{2}{c}{(2--10~keV)} 	 & \multicolumn{2}{c}{(0.5--10~keV)}  		& 				\\  
\hline
00010563002 & 2018-02-21	& 0.97 	& $1.53 \pm 0.04$ 	& $5.9 \pm 0.2$		&  $4.5\pm 0.1$	& $8.4 \pm 0.2$  	 	&	$6.4\pm 0.1$		& 1.00 (288) \\
00010563003 & 2018-02-23	& 0.83 	& $1.48 \pm 0.05$ 	& $7.1 \pm 0.3$ 		&  $5.5\pm 0.2$	& $9.9 \pm 0.3$ 			&	$7.6\pm 0.2$		& 1.33 (174) \\
00010563004 & 2018-02-25	& 0.41 	& $1.63 \pm 0.05$ 	& $6.3 \pm 0.3$ 		&  $4.8\pm 0.2$	& $9.4 \pm 0.3$ 			&	$7.2\pm 0.2$ 	& 0.94 (189) \\
00088654001 & 2018-02-27	& 1.8 	& $1.53 \pm 0.02$ 	& $5.7 \pm 0.1$ 		&  $4.4\pm 0.1$	& $8.1 \pm 0.1$ 			&	$6.2\pm 0.1$ 	& 1.13 (496) \\
00010563005 & 2018-03-01	& 0.48 	& $1.50 \pm 0.05$ 	& $5.1 \pm 0.2$ 		&  $3.9\pm 0.2$	& $7.1 \pm 0.2$ 			&   $5.5\pm 0.2$		& 1.27 (165) \\ 
00010563006 & 2018-03-02	& 0.90 	& $1.57 \pm 0.04$ 	& $5.0 \pm 0.2$ 		&  $3.8\pm 0.1$	& $7.2 \pm 0.1$ 			&	$5.5\pm 0.1$		& 1.10 (286) \\
00010563007 & 2018-03-05	& 0.93 	& $1.58 \pm 0.04$ 	& $4.7 \pm 0.1$ 		&  $3.6\pm 0.1$	& $6.9 \pm 0.1$ 			&	$4.0\pm 0.1$		& 0.99 (282) \\
00088654002 & 2018-03-06	& 1.8  	& $1.54 \pm 0.03$ 	& $4.5 \pm 0.1$ 		&  $3.4\pm 0.1$	& $6.4 \pm 0.1$ 			&	$4.9\pm 0.1$		& 1.01 (414) \\
00088654003 & 2018-03-07	& 2.24  	& $1.51 \pm 0.03$ 	& $4.5 \pm 0.1$ 		&  $3.5\pm 0.1$	& $6.4 \pm 0.1$ 			&	$4.9\pm 0.1$		& 1.00 (468) \\
00010563008 & 2018-03-17	& 0.77 	& $1.53 \pm 0.05$ 	& $3.7 \pm 0.1$ 		&  $2.8\pm 0.1$	& $5.2 \pm 0.1$ 			&	$4.0\pm 0.1$		& 0.96 (191) \\
00088654004 & 2018-03-25	& 1.48	& $1.67 \pm 0.03$ 	& $4.8 \pm 0.1$ 		&  $3.6\pm 0.1$	& $7.3 \pm 0.2$ 			&	$5.6\pm 0.1$		& 0.98 (396) \\

\hline
\end{tabular}
\label{tab:swiftres}
\begin{tablenotes}
\item[] $^{\rm a}$ Unabsorbed fluxes in units of $10^{-10}~\flux$.
\item[] $^{\rm b}$ Luminosities in units of  $10^{36}$ $\lum$. 
\end{tablenotes}
\end{threeparttable}
\end{table*}


\begin{table*}
\centering
\caption{Fitting results for the  \xmm\ observation. Uncertainties are expressed at 90 per cent confidence level.}
\begin{threeparttable}
\begin{tabular}{ l c  c c}
\hline
Component /Model												& 	DISKBB+NTHCOMP+GAUSS	 		&	BB+NTHCOMP+GAUSS	 				&   DISKBB+COMPPS+GAUSS\\		 			
\hline
\Nh\ ($\times 10^{22}$ \nh)									& 	1.08$\pm$0.07 				&	1.03$\pm$0.05					&	1.14$\pm$0.07				\\
$kT_{\rm in}$ (keV)											& 	0.23$\pm$0.03			 	&	0.18$\pm$0.01					&	0.21$\pm$0.03\\
$N_{\rm in/bb}$/	$R_{\rm in/bb}$ (km)							& 	1222$^{+2686}_{-709}$/ 80$^{+63}_{-29}$&	2915$^{+3197}_{-1596}$ / 43$^{+19}_{-14}$	&	2633$^{+5522}_{-1902}$ / 117$^{+89}_{-56}$\\
$\Gamma$	/y			 										& 	1.55$^{+0.04}_{-0.01}$	    &  	1.54$\pm$0.02   					&  	1.5$\pm$0.1   \\
$kT_{\rm e}$ (keV)											& 	50 (fix)						& 	50(fix)		 					& 	50(fix)		 		\\
Refl															&    --							&   --                               &   0.31$^{+0.34}_{-0.25}$               \\
$N_{\rm nthcomp/compps}$ ($\times 10^{-2}$) 					& 	6.14$\pm$0.4					&	5.8$\pm$0.3 						&	3537$^{+2334}_{-1586}$ 	\\
$E_{\rm gaus}$ (keV)											&	6.7$^{+0.4}_{-0.2}$			&	6.7	$^{+0.3}_{-0.2}$				 &	7.1	$^{+0.4}_{-0.2}$\\
$\sigma_{\rm gaus}$ (keV)									& 	0.68$^{+1.7}_{-0.2}$			&	0.67$^{+0.7}_{-0.3}$				&	1.16$^{+1.3}_{-0.4}$\\
$k_{\rm gaus}$ ($\times 10^{-4}$ photons cm$^{-2}$ s$^{-1}$) 	& 	2.5$^{+7}_{-1}$				&	2.5$^{+4}_{-1}$   				&	3.9$^{+15}_{-2}$\\
\qhis\ (dof) 												& 	541.81 (505)					&	546.83 (505)			  			&	542.72 (504)	\\
 															&								&									&					\\
\multicolumn{1}{r}{(0.5--10 keV)} \\
$F_{\rm X, unabs}$ ($\times10^{-10} \flux$)					&	5.1$\pm$0.2					&	4.87$\pm$0.1						&	5.3$\pm$0.2		 	\\
\lx$^{\rm a}$ ($\times10^{36}\lum$)							&	3.9$\pm$0.2					&	3.73			 					&	4.0$\pm$0.2			\\
Thermal fraction$^{\rm b}$ ($\%$)							&	8							&	5 				 				&     8						\\

 															&								&					\\
\multicolumn{1}{r}{(2--10 keV)} 								&								&					\\
$F_{\rm X, unabs}$ ($\times10^{-10} \flux$)	 				&	3.40$\pm$0.01				&	3.38	$\pm$0.01 					&	3.4$\pm$0.02		\\
\lx$^{\rm a}$ ($\times10^{36}\lum$)							&	2.6$\pm$	0.00	7				&	2.59	$\pm$	0.00	7				&	2.6	$\pm$0.02		\\
Thermal fraction$^{\rm c}$ ($\%$)			   				&	1							&	< 1 								&	2				\\

\hline
\end{tabular}
\begin{tablenotes}
\item[a]{Unabsorbed luminosity assuming a distance of 8~kpc.}
\item[b]{Fractional contribution of the thermal component to the total 0.5--10~keV flux.}
\item[c]{Fractional contribution of the thermal component to the total 2--10~keV flux.}

\end{tablenotes}
\label{tab:xmmres}
\end{threeparttable}
\end{table*}

\subsection{X-ray analysis}\label{sub:resXray}
We used \textsc{xspec} (v.12.9.1; \citealt{Arnaud1996}) to analyse the X-ray spectra. In order to account for interstellar absorption, we used the Tuebingen-Boulder Interstellar Medium absorption model (TBABS in \textsc{xspec}) with cross-sections of \citet{Verner1996} and abundances of \citet{Wilms2000}. In this work we assume a distance of 8~kpc \citep{Russell2018} and an orbital inclination of 60\degr.

\subsubsection{\xmm\ temporal analysis}\label{subsub:temp}
EPIC PN light curves in the energy band 1-10 keV were extracted with a time bin of 6 ms. We computed the power spectral density function (PSD) by averaging estimates obtained from segments of 60~s length 
. The Poisson noise level was estimated by fitting a constant model at frequencies $>30$~Hz, where variability associated with counting noise dominates. Fig. \ref{fig:pow_spec} shows the resulting Poisson noise-subtracted PSD, with fractional root-mean-square (rms) normalization \citep{Miyamoto1991}. A flat-topped broad band noise distribution is observed, which is typical of X-ray binaries in the hard state. Moreover, we find no evidence of quasi-periodic oscillations (QPOs). This is also true if we increase the signal-to-noise ratio by using shorter segments (e.g. 16s) when computing the PSD. In both BH and NS LXRBs, strong, low-frequency QPOs are typically observed in high luminosity states, mostly close to, or during the transition to/from the soft state  \citep[e.g.][]{Motta2017}. Thus, the lack of QPOs is indicative of a low luminosity hard state. 
Moreover, the integrated 0.1--50~Hz fractional rms yields $31 \pm 1$ per cent, again typical of low luminosity hard states \citep{Munoz-Darias2011a, Munoz-Darias2014a}. 

\subsubsection{\xmm\ spectral analysis}\label{subsub:spec}

We simultaneously fitted the  0.6--1.8~keV RGS spectra (RGS1 and RGS2 first order) and the 0.7--10~keV EPIC-PN spectrum with the parameters tied between the three detectors. We added a constant factor (CONSTANT) to the spectral models with a value fixed to 1 for EPIC--PN spectrum and free to vary for the RGS spectra to account for cross-calibration uncertainties between the instruments. We added a 1 per cent systematic error to all instruments to account for uncertainties in the relative calibration between the RGS and EPIC--PN detectors (\citealt{Kirsch2004}; XMM-SOC-CAL-TN-0052). 

In a first attempt, we fitted the spectra with a single absorbed thermally Comptonized continuum model (NTHCOMP in \textsc{xspec}; \citealt{Zdziarski1996, Zycki1999}), which is parametrised by an asymptotic power law index ($\Gamma$), the corona electron temperature ($kT_{\rm e}$) and the up-scattered seed photons temperature ($kT_{\rm seed}$). The $kT_{\rm e}$ temperature generally adopts values higher than 10~keV in the hard state, which is beyond our spectral coverage (0.5--10~keV). Hence, we fixed this parameter to the reported e-folding energy of exponential rolloff obtained with \inte\ spectra \citep{Fuerst2018}. This is $kT_{\rm e}$=50~keV assuming a relation  $E_{\rm c}=140~keV\approx3kT_{\rm e}$  \footnote{The actual value of $kT_{\rm e}$ is within 30--90~keV (since $E_{\rm c}=140^{+120}_{-50}$~keV).  Fixing it to 50~keV does not affect significantly the resulting spectral parameters.}.  However, with a \qhis\ of 636.9 for 509 degrees of freedom (dof) the model returned a non acceptable fit (p-value\footnote{The probability value (p-value) represents the probability that the deviations between the data and the model are due to chance alone. In general, a model can be rejected when the p-value is smaller than 0.05.} of $9\times10^{-5}$). In addition to a soft excess, the fit residuals showed two emission features at $\sim$1.8 and $\sim$2.2~keV produced, most likely, by the detector silicon K-edge (1.8~keV) and the mirror AuM-edge (2.3~keV) \citep{Guainazzi2014}. Therefore, from now on, we added two Gaussian components (GAUSS) in all our models to mitigate these instrumental features. Their inclusion to our model did not significantly improve the fit  ($\Delta$\qhis=4). 

In a second step, we added to our previous model a thermal soft component, either a multicolour disc (i.e, DISKBB+NTHCOMP; \citealt{Mitsuda1984,Makishima1986}) to account for emission from the accretion disc or a single blackbody (i.e, BBODYRAD+NTHCOMP) to account for the emission from the surface/boundary layer in the case of a NS. We assumed that the seed photons arise from the corresponding thermal component. Thus, we coupled $kT_{\rm seed}$ to either $kT_{\rm in}$ or $kT_{\rm bb}$, respectively, and changed the seed photons shape parameter accordingly (im-type in NTHCOMP). Both models improved significantly the fit (\qhis$\cong$577 for 508 dof, F-test probability $\sim10^{-12}$), although they still did not reproduce adequately the data (p-value$\cong$0.01). Both fits leave residuals in the region of $\sim$6.4~keV consistent with Fe K$\alpha$ emission (see lower panel in Fig.\ref{fig:spectra}). Therefore, we included a Gaussian component (GAUSS), which resulted in an acceptable fit (p-value$\cong$0.1). The line converged to a central energy of $E_{\rm l}\sim$6.7~keV and $\sigma_{\rm l}\sim$0.7~keV with a equivalent width (EW) of 75~eV. When assuming soft emission arising from the accretion disc we got an equivalent hydrogen column (\Nh) of ($1.08\pm0.07)\times10^{22}$~\nh. The obtained temperature at the inner disc radius ($kT_{\rm in}$) is $0.23\pm0.03$~keV. The normalization can be translated to an apparent inner disc radius of $R_{\rm in}$ $\sim$80~km (with \citealt{Kubota1998} correction applied). The Comptonization asymptotic power-law photon index ($\Gamma$) is $1.55\pm0.04$. The inferred 0.5--10~keV unabsorbed flux is (5.1$\pm$0.2)$\times10^{-10}~\flux$, to which the thermal component contributes 8~per cent. Likewise, in the 2--10~keV band, we measure (3.40$\pm$0.01)$\times10^{-10}~\flux$ and a thermal contribution of 1 per cent. Similar values are recovered for the Comptonizing parameters when using a single blackbody as the soft thermal component (see Table \ref{tab:xmmres}). The resulting blackbody temperature was $kT_{\rm bb}$=$0.18\pm0.01$~keV with an inferred emission radius of $R_{\rm bb}$=43$^{+19}_{-14}$~km.

We also tried the three-component model (DISKBB+BBODYRAD+NTHCOMP+GAUSS) proposed for NS hard state spectra when high-quality coverage at low energies is available (\citealt{ArmasPadilla2017,ArmasPadilla2018}; see also \citealt{Lin2007}). The model adequately fits the spectra (\qhis$\cong$540 for 503 dof), but the extra thermal component is not statistically required (F-test probability > 0.1).

With the aim of comparing with literature, in particular with the work of \citet{Burke2017}, we also modelled the Comptonised emission using the COMPPS model \citep{Poutanen1996}. Following this work  we used the model DISKBB+COMPPS+GAUSS with the same parameter constraints (section 3 and 5.5 of \citealt{Burke2017}). This model returned consistent values for \Nh, the thermal component  and the fluxes (see Table \ref{tab:xmmres}). With regards to the Comptinzing component,  the y-parameter (y) obtained is $\sim$1.5 and the reflection strength (R=$\Omega/2\pi$) $\sim$0.3. The resulting line central energy is slightly higher  ($E_{\rm l}\sim$7.1~keV and $\sigma_{\rm l}\sim$1.2~keV) than that derived above, probably due to coupling with the Fe edge (7.1~keV) of the reflection continuum. In fact, if we fix the reflection strength to zero, the Gaussian parameters converge to similar values than those obtained with NTHCOMP.

\subsubsection{Spectral-timing analysis}\label{subsub:fourier}

In the last two decades significant efforts have been dedicated to study the effects of the different spectral components in the time domain (e.g. \citealt{Gilfanov2000}, \citealt{Yamada2013}). With this aim, we computed the fractional-rms as a function of the energy (rms spectrum) using the \xmm\ PN data and following for every energy bin the procedure described in Sec \ref{subsub:temp}. This resulted in a flat rms spectrum, which is typical of BHs in the hard state (see e.g. fig. 4 in \citealt{Belloni2011}). Subsequently, we used this rms spectrum to rescale the energy spectrum, which yielded the rms-scaled energy spectrum \citep[e.g.][]{Uttley2014}. It shows the spectral shape of the components contributing to the observed variability over a given time scale (0.1-50 Hz in this case). Given the shape of the PSD (Fig. \ref{fig:pow_spec}) this frequency band encompasses almost all the observed X-ray variability. The rms-scaled spectrum is shown in Fig. \ref{fig:RMS_spec} together with the data-to-model ratio obtained after fitting it with a thermal comptonization model [TBABS*NTHCOMP in Xspec with \Nh\ fixed to that derived from the regular spectral fits (Sect. \ref{subsub:spec})]. The fit reveals clear excess residuals (bottom panel) at low energies (E$\lesssim$2~keV), suggesting the presence of significant variability associated with the soft component (i.e. the DISKBB component; see Fig. \ref{fig:spectra}). This residual is also significantly detected when splitting the variability over long (0.1--1 Hz) and short (1-50 Hz) time scales. 

We note that additional residuals are seen at the energy of the Fe line. Adding a Gaussian component yields a centroid energy of $\sim 6.6$~keV, consistent with the laboratory energy of the line. However, this feature is not significant in the rms-scaled spectrum ($\Delta$\qhis=2 for 3 dof).

\subsubsection{\swift\ analysis}\label{subsub:swiftres}

All \swift\ spectra are well described by an absorbed power-law model, in which we fixed the \Nh\ to the value obtained from the \xmm\ fit (\Nh=$1.1\times10^{22}$~\nh). Results are reported in Table \ref{tab:swiftres} and Fig \ref{fig:swift}. Across the 34~d \swift/XRT monitoring, $\Gamma$ fluctuated in the range of $\sim$1.45--1.70 and the 0.5--10~keV unabsorbed flux varied by a factor of 2. The flux peaked on February 23 with a value of (9.9$\pm$0.3)$\times 10^{-10}~\flux$ and smoothly decreased to a (5.2$\pm$0.2)$\times 10^{-10}~\flux$ over the following 27 days. On March 25 there was a small re-brightening after which the source returned to quiescence (see Fig \ref{fig:LC}). Assuming a distance of 8~kpc, the above correspond to a luminosity range of (4 -- 7.6) $\times 10^{36}~\lum$. The Harness Ratio (HR; ratio of counts in the 2--10~keV and 0.5--2~keV energy bands) oscillates in the range of $\sim$1.4--1.6. These values indicate that the source remained in the hard state during the full outburst and are fully consistent with the \xmm\ timing and spectral results. 

\begin{figure}
\begin{center}
\includegraphics[keepaspectratio,width=\columnwidth, trim=0.0cm 0.0cm 0.0cm 0.0cm]{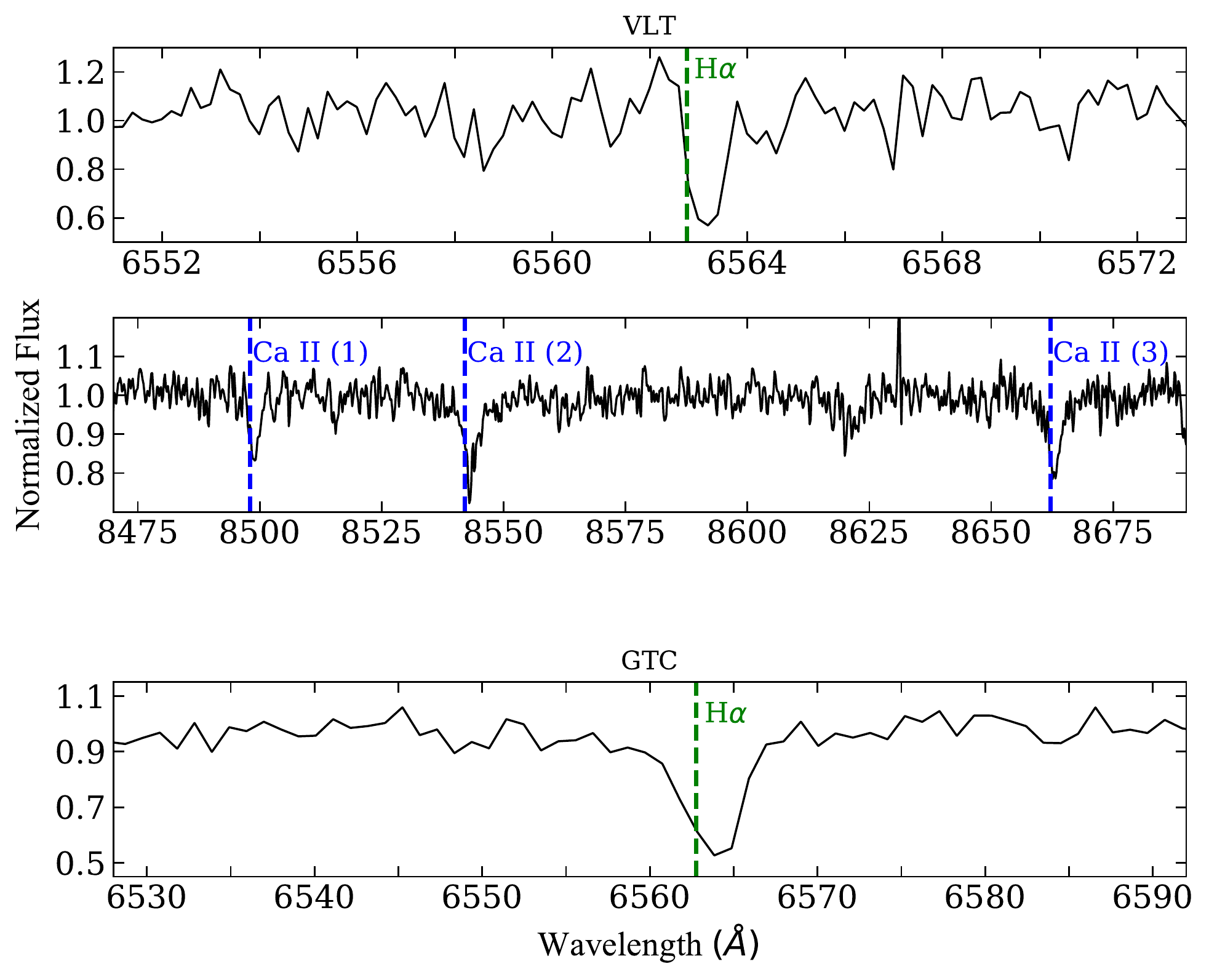}\\
\caption{Two top panels: VLT/X-shooter spectrum taken during outburst. The H$_{\alpha}$ (green) and CaII Triplet (blue) absorptions are marked with dotted lines. Bottom panel: GTC/OSIRIS spectrum in the H$_{\alpha}$ region taken during quiescence.}
\label{fig:Opt}
\end{center}
\end{figure}

\subsection{Optical spectroscopy}\label{sec:Star}
The outburst VLT spectrum is shown in the upper panels of Fig. \ref{fig:Opt}. The strongest features detected correspond to the \ion{Ca}{ii} triplet at 8500--8600 \AA. These absorption lines do not typically arise from accretion discs but from stars with spectral types later than F. However, in M stars these are accompanied by very broad absorption components of different elements not present in our data, while early F stars show also weak Paschen transitions on the red wing of the \ion{Ca}{ii} absorptions. In addition, we detect weaker absorption of \ion{Mg}{i} at  8806.8 \AA\ (not shown), also typical of late spectral type stars. All the above suggests that the outburst optical spectrum is dominated by the emission from a G--K star.

Unfortunately, the quiescent spectrum (GTC) did not cover the \ion{Ca}{ii} triplet region. Instead, we clearly detected \ha\ in absorption with an equivalent with of $2.7 \pm 0.2$~\AA. A Gaussian fit to this line reveals that it is shifted by $40 \pm 5$ \kms\ with respect to the laboratory wavelength. The same technique applied to the \ion{Ca}{ii} triplet (i.e. multi-Gaussian fit) and \ion{Mg}{i} detected in outburst yields a velocity offset of $36 \pm 1$~\kms, that is, consistent with the quiescence value. Unless both observations were taken (by chance) at the same orbital phase or the orbital inclination is (very) close to zero, the above suggests that the absorption lines arise from a field, interloper star (i.e. not from the donor star).

A careful inspection of the outburst spectrum also reveals a weak \ha\ absorption, in this case with an equivalent width of $0.33 \pm 0.05$ \AA\ (see top panel in Fig \ref{fig:Opt}). The line is centred at +17~\kms, slightly (but significantly) different than both the \ion{Ca}{ii} triplet velocity derived from the same spectrum and that of \ha\ itself measured in quiescence.  This can be explained if the emission from the interloper is veiled by the actual optical emission from the accretion disc during outburst. Indeed, this extra emission must be present in order to account for the $\sim 0.8$ mag brightening that we determined from the acquisition images. It is also expected to include additional \ha\ emission (but not \ion{Ca}{ii}), since this is the strongest emission line typically detected in LMXBs in outburst \citep{MataSanchez2018}. Interestingly, a rise of $\sim$ 0.8 mag in the continuum from quiescence to outburst, would imply a drop in the equivalent width by a factor of 2.1 (i.e., 2.7/2.1= 1.3 \AA\ would be expected in outburst). Since we measure an equivalent width of $\sim 0.3$ \AA\ this suggests that the absorption is indeed filled by extra emission. The latter might account for the different velocity offset as it is likely centred at a different systemic velocity than that of the interloper.       

\section{Discussion}\label{sec:Disc}

In this work, we report on the X-ray and optical spectroscopic observations as well as the long--term evolution of \MAXI\ during its discovery outburst. The X-ray spectral properties, the outburst intensity and duration, and the temporal properties, indicate that \MAXI\ is a new member of the family of LMXBs. 

According to the \swift/BAT light curve (Fig. \ref{fig:LC}) the duration of the outburst was $\sim$75~d with the first detection on  February 19, 2018 (we notice that the source was not observable before February 14 due to Sun-constraints). If we assume a distance of 8~kpc, we obtain a 0.5--10~keV peak luminosity of $7.6\times10^{36}~\lum$. The integrated fractional rms inferred from the \xmm\ data is $\sim$31~per cent (0.1--50 Hz), which indicates that the source was in the hard state at that time \citep{Munoz-Darias2011a,Munoz-Darias2014a}. Furthermore, the photon index values are in the range of $\sim$1.4--1.7 along the whole outburst, which strongly suggests that the source never transited to the soft state \citep{McClintock2006,Done2007} and remained in the hard state during the event. This behaviour is not unheard in the family of LMXBs. Indeed, sources always showing only hard state outburst as well as systems typically displaying regular outburst with occasional (typically fainter) only hard state events (a.k.a. failed outburst) have been observed \citep[e.g.][]{Capitanio2009, Motta2010, ArmasPadilla2013a, Koljonen2016,Munoz-Darias2014a}.

In the absence of dynamical measurements, thermonuclear X-ray bursts and coherent pulsations, the spectral characteristics can help to infer the nature of the compact object. However, we have been able to model the \xmm\ spectra by using both BH-like and NS-like models. Therefore, we have to scrutinize the obtained parameters in order to make an attempt to tag the accretor class. Two of the most revealing model parameters  are related to the Comptonizing Corona. These are the electron temperature and optical depth, which combine in the so-called Compton y-parameter\footnote{Compton y-parameter is defined as $y=\frac{4kT}{mc^{2}}\rm Max(\tau,\tau^{2})$, where $\tau$ is the optical depth.}. This was best showed in \citet{Burke2017}, who found a dichotomy between the Comptonization properties of BH and NS systems while in the hard state. In particular, they found a boundary at y$\approx$0.9, with NSs sitting below this value and BHs above it. The difference implies that in a Corona with a given optical depth, the temperature will be lower for NS systems, which supports the idea that the properties of the Comptonizing Corona are affected by the additional seed photons coming from the NS \citep{Sunyaev1989}. In fact, \citet{Burke2017} found that while the electron temperature $kT_{\rm e}$ in BH systems can cover a broad range, from $kT_{\rm e}$ $\sim$ 30 to 200 keV, $kT_{\rm e}$ peaks at $\sim$15--25 keV for NSs, with very few spectra exceeding $kT_{\rm e}$ $\sim$50--70~keV. In the case of \MAXI, the electron temperature is in the rage $kT_{\rm e}\sim$30--90~keV (as inferred from the high-energy turn-off reported by \citealt{Fuerst2018}) and the Compton y-parameter is $\sim$1.5 (see Section \ref{subsec:xmm}). Although the Corona temperature is fully compatible with that displayed by BH systems, this argument is not conclusive, since NSs can also reach such high values. The y-parameter, on the other hand, is well above the 0.9 boundary clearly favouring a BH accretor. Our y-value implies a Corona with an optical depth $\tau\sim$4 ($kT_{\rm e}\sim$50~keV) for \MAXI, while an optical depth of $\sim$2 is expected for NS systems at similar electron temperatures. In addition, it has also been shown than the equivalent width of the Fe reflection line is anti-correlated with the y-parameter \citep{Gilfanov1999,Burke2017}. It is worth mentioning that we measure an equivalent width of $\sim$75~eV, which nicely sits in this relation when considering y-parameter $\sim$1.5 .

The spectral parameters derived from the thermal component also suggest, although to a lesser extent, a BH accretor. The temperature ($\sim$0.2~keV) and the normalization (> 10$^{3}$) of the disc component, which contributes less than 10~per cent to the 0.5--10~keV flux, are fully consistent with values reported for BHs in the hard state \citep[e.g.][]{Reis2010,ArmasPadilla2014a,Shidatsu2014,Plant2015}. NS systems accreting at low luminosities generally display hotter blackbody temperatures ($kT_{\rm bb}\ga0.3$~keV) and lower normalizations (< 10$^{3}$), with this component typically contributing at the $\sim$20-50~per~cent level \citep{Degenaar2017,Sharma2018, DiSalvo2015}. However, this fraction is anti-correlated with luminosity, and can be as low as $\la$0--10~per~cent at high luminosities (\lx$\ga10^{36}~\lum$; \citealt{Campana2014,Allen2015}). Nevertheless, this lower thermal contribution would be accompanied by an increase on the black body temperature, from  $kT_{\rm bb}\sim$0.3--0.5~keV at \lx$\la10^{35}~\lum$ to $kT_{\rm bb}\sim$1--2~keV at \lx$\ga10^{36}~\lum$ \citep{Lin2007,Allen2015}. Finally, our analysis of the rms-scaled spectrum suggests that in addition to the Comptonization component, the disc also contributes to the observed variability within the frequency range of 0.1--50 Hz. Disc variability has been reported in several BHXRBs \citep[e.g.][]{Wilkinson2009,DeMarco2015} and seems to be a characteristic feature of low luminosity hard states. All in all, the X-ray spectral characteristics of \MAXI\ clearly favour a BH accretor. 

\subsection{On the optical counterpart and the distance to the source}
Unfortunately, our optical spectra do not shed much light on the binary properties of \MAXI. We find that both outburst and quiescence data show stellar absorption features instead of broad emission lines. In Section \ref{sec:Star}, we explain that the absence of velocity shifts between outburst and quiescence, together with the evolution of the equivalent width of \ha, strongly suggest that the optical emission from the system is dominated by that of an interloper star. This odd behaviour resembles that observed in other LMXBs, such as the prototypical NS transient Aql X-1, although in this case the accretion disc outshines the interloper in outburst \citep[e.g.][and references therein]{MataSanchez2017}.  
The faintness of the source (i=19.7 in quiescence) and the presence of strong \ion{Ca}{ii} features in absence of H Paschen components, clearly favour a spectral type G or K. Additionally, the equivalent width of the \ha\ absorption line is correlated with the spectral type for cool, non-active stars (F to mid K; \citealt{Scholz2007}). The value derived from our quiescence data (2.7 \AA) points towards a $\sim$ G5 star. In addition to the PS1 catalog, an infrared counterpart to \MAXI\ is present in the UKIDSS-DR6 Galactic Plane Survey \citep{Lucas2008}. The object is K=16.0 and its coordinates accurately match those of the optical counterpart.  Adopting a G5-V spectral type we derive a distance of 3.4 kpc to the interloper star. To obtain this estimation we have neglected extinction effects and used the K absolute magnitude from \citet{Pecaut2013}. 

Companion stars in LMXBs  have typically G--K spectral types but these stars would not dominate the optical emission from the X-ray transient in outburst (e.g. \citealt{Charles2006}). Even in the very atypical case of the BH transient V4641 Sgr with a $\sim$ B9 donor, a strong \ha\ emission component is detected in many spectra (\citealt{Munoz-Darias2018} for a recent study). On the other hand, the presence of an interloper star naturally explains the very odd outburst amplitude of only $\sim$ 1 mag (see \citealt{Corral-Santana2016} for typical outburst amplitudes in BH systems).  Given the late spectral type of the proposed interloper and the fact that LMXB accretion discs outshine their companions (of similar spectral types) by several orders of magnitude during outburst, we conclude that \MAXI\ is most likely located further away than the interloper at a distance $\gtrsim 3$~kpc.

\section*{Acknowledgements}

We thank the anonymous referee for constructive comments that improved the quality of this paper. We acknowledge support by the Spanish MINECO under grant
AYA2017-83216-P. MAP's research is funded under the Juan de la Cierva Fellowship Programme (IJCI-2016-30867). TMD and MAPT acknowledge support via a Ramon y Cajal Fellowship (RYC-2015-18148 and RYC-2015-17854, respectively).BDM acknowledges the Polish National Science Center for support under grant Polonez 2016/21/P/ST9/04025 and the European Commission for support under H2020-MSCA-IF-2017 action, Grant. No. 798726 BHmapping. We acknowledge the use of public data from the Swift data archive. \xmm\ is an ESA science mission with instruments and contributions directly funded by ESA Member States and NASA. Based on observations made with the Gran Telescopio Canarias (GTC), instaled in the Spanish Observatorio del Roque de los Muchachos of the Instituto de Astrof\'isica de Canarias, in the island of La Palma. Based on observations collected at the European Southern Observatory under ESO programme 0100.D--0292(A). The authors are thankful to the GTC and VLT team that carried out the ToO observations. MOLLY software developed by T. R. Marsh is gratefully acknowledged.



\bibliographystyle{mnras}
\bibliography{MAXIJ1813.bbl} 



%
%


\bsp	
\label{lastpage}
\end{document}